\documentclass[twocolumn]{autart_arxiv}    

\usepackage{graphicx}          
\usepackage{wrapfig}
\usepackage{amsmath} 
\usepackage{amssymb}  
\usepackage[dvipsnames]{xcolor}
\usepackage{url}
\usepackage{natbib}

\newtheorem{corollary}{Corollary}
\newtheorem{lemma}{Lemma}
\newtheorem{assumption}{Assumption}
\newtheorem{remark}{Remark}
\newtheorem{definition}{Definition}
\newtheorem{problem}{Problem}

\begin{document}
	
	\begin{frontmatter}
		\title{Formation control of a leader-follower structure in three dimensional space using bearing measurements\thanksref{footnoteinfo}} 
		
		\thanks[footnoteinfo]{This paper was partially presented at the 21st IFAC World
			Congress, 2020. Corresponding author Zhiqi Tang. }
			\thanks[footnoteinfo1]{Carlos Silvestre is on leave from the Instituto Superior Técnico of the Universidade de Lisboa, Portugal}
		\author[lisbon,Nice]{Zhiqi Tang}\ead{zhiqitang@tecnico.ulisboa.pt},    
		\author[lisbon]{Rita Cunha}\ead{rita@isr.tecnico.ulisboa.pt},               
		\author[Nice,i3s]{Tarek Hamel}\ead{ thamel@i3s.unice.fr},
		\author[macau]{Carlos Silvestre\thanksref{footnoteinfo1}}\ead{ csilvestre@umac.mo}
		\address[lisbon]{ISR, Instituto Superior T\'{e}cnico, Universidade de Lisboa, Portugal. }  
		\address[Nice]{I3S-CNRS, Universit\'{e} C\^{o}te d'Azur, Nice-Sophia Antipolis, France.}             
		\address[i3s]{IUF, Institut Universitaire de France, Nice-Sophia Antipolis, France.}        
		\address[macau]{Faculty of Science and Technology of the University of Macau, Macao, China.}
		
		\begin{abstract}                          
			This paper addresses the problem of bearing leader-follower formation control in three-dimensional space by exploring the persistence of excitation (PE) of the desired formation. Using only bearing and relative velocity measurements, distributed control laws are derived for a group of agents with double-integrator dynamics. The key contribution is that the exponential stabilization of the actual formation to the desired one in terms of both shape and scale is guaranteed as long as the PE conditions on the desired formation are satisfied. The approach generalizes stability results provided in prior work for leader-first follower (LFF) structures which are based on bearing rigidity and constraint consistency to ensure the exponential stabilization of the actual formation to a desired static geometric pattern up to a scale factor. Simulations results are provided to illustrate the performance of the proposed control method.
		\end{abstract}
	\end{frontmatter}
	\section{Introduction}
	The formation control problem has been extensively studied over the last decades both by the robotics and the control communities. The main categories of solutions can be classified as follows (\cite{oh2015survey}): \emph{i)}  position-based formation control, \cite{ren2007distributed},  \emph{ii)} displacement-based formation control, \cite{ren2005coordination}, \emph{iii)} distance-based formation control, \cite{anderson2007control}, and more recently \emph{iv)} bearing-based formation control, \cite{basiri2010distributed}. This latter category has received growing attention due to its minimal requirements on the sensing ability of each agent. Early works on bearing-based formation control were limited to planar formations and were  mainly focused on controlling the subtended bearing angles which are measured in each agent's local coordinate frame (see \cite{basiri2010distributed} and \cite{bishop2011very}). The main body of work however builds on concepts of bearing rigidity theory, which investigates the conditions for which a static geometric pattern of a formation is uniquely determined by the corresponding bearing measurements.  Bearing rigidity theory in two-dimensional space (also termed parallel rigidity) is explored in \cite{eren2003sensor} and \cite{servatius1999constraining}.
	More recently, it has been extended to an arbitrary dimensional space along with a formation control solution based on bearing measurements in \cite{zhao2016bearing}, when the graph is undirected. Under the assumption that the desired formation is infinitesimally bearing rigid, the resulting bearing controller guarantees convergence to the target formation up to a scaling factor and  translation vector.	In the more challenging context of directed graphs, achieving stabilization of a formation requires not only bearing rigidity, as in the case of undirected graphs, but also constraint consistence, which is the ability to maintain consistence between constraints induced by the desired bearing measurements (termed bearing persistence, in  \cite{zhao2015bearing}). In \cite{eren2012formation}, the conditions for directed bearing rigidity of a digraph in two-dimensional space are stated and a bearing control law for nonholomonic agents is proposed. In \cite{trinh2019bearing}, bearing control laws that asymptotically stabilize leader-first follower (LFF) formations to the desired formations up to a translation (the leader's position) and a scaling factor have been proposed.
	Since bearing rigidity of a static formation is invariant to scale, the measure of at least one distance between two agent is required to guarantee the convergence of formations in terms of shape and scale. For instance, in \cite{schiano2016rigidity} a controller based on bearing rigidity of directed bearing frameworks defined in $\mathbb{R}^3 \times \mathcal{S}^1$ complemented with the measure of at least one distance between two agent is proposed. 
	
	\begin{figure}[!t]
	\centering
		\includegraphics[width=3.4in]{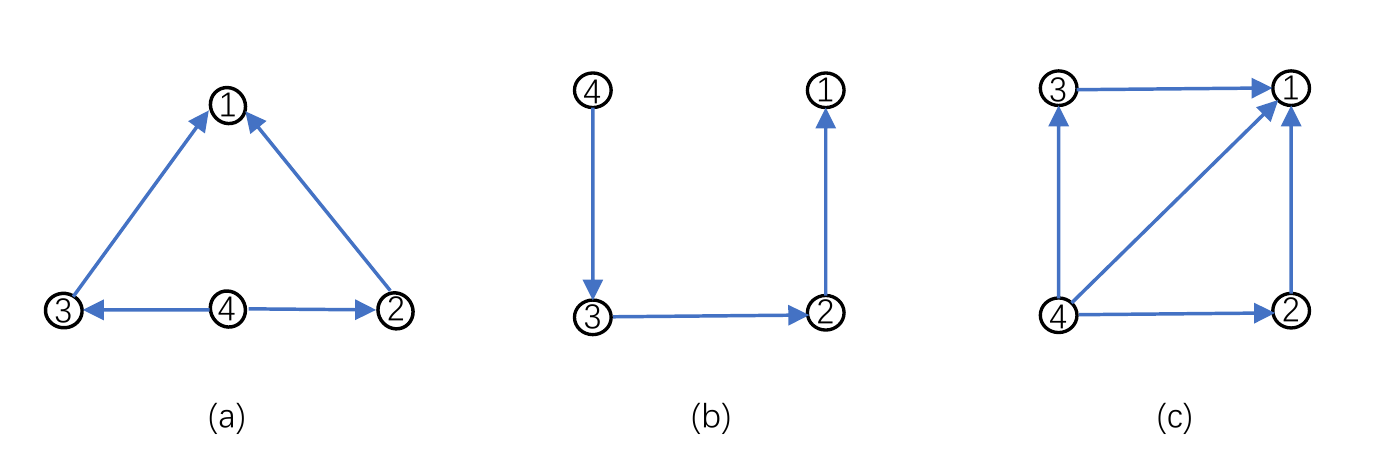}	
		\caption{Examples of leader-follower formations. The formations in (a) and (b) are not bearing rigid and in (c) is bearing rigid but not constraint consistent. The asymptotic stability of these three formations can not be guaranteed using bearing controllers relying only on bearing rigidity theory and constraint consistence (\cite{zhao2015bearing}). It is however guaranteed in this paper under the proposed PE condition.}
		\label{fig:notrigid}
	\end{figure}
	In this paper, we consider the problem of controlling a leader-follower formation (i.e. a formation under a directed acyclic graph that has a spanning tree, see Fig.~\ref{fig:notrigid}) using only bearing and relative velocity measurements. We particularly focus on the problem of stabilizing the formation's geometric pattern to a desired one by exploiting persistence of excitation (PE) of the bearings of the desired formation. Note that this PE condition can be enforced a priori and has no dependence on the initial conditions of system.
The concept of persistence of excitation (PE) is a well-known concept in adaptive control and identification of linear systems. It has been recently exploited for position estimation from bearing and biased velocity measures in \cite{le2017observers}, \cite{hamel2017position}. This paper generalizes prior work reported in \cite{tang2020bearing}, that proposes a kinematic bearing leader-follower formation control law. The main distinctions with respect to \cite{tang2020bearing} are: \emph{i)} the explicit treatment of leader-follower formation with double-integrator dynamics and, \emph{ii)} the introduction of a generalised rigidity concept: \textit{relaxed bearing rigidity}, which makes the connection between bearing PE and bearing rigidity theory.
	The key contribution is to show that the required classical conditions on the graph topology (bearing rigidity and constraint consistence) used to guarantee stabilization of the formation to a desired shape up to a scale are relaxed here in a natural manner by exploiting PE of the bearing information generated by the desired formation. The proposed control approach draws inspiration from the work in \cite{trinh2019bearing}, which presents a first-order bearing formation control law, considering a LFF graph topology. A distinctive feature of the present work is the shift of focus from static formations to time-varying formations. The approach relies on the simplicity of controllers that guarantee exponential stability of the formation towards the desired one in terms of shape and scale when the bearing PE conditions are fulfilled.
	
	The body of the paper is organized as follows. Section \ref{sec:prelmn} presents mathematical background on graph theory and introduces the definition of bearing PE together with conditions for bearing PE. Section \ref{sec:persistent} describes the concepts of bearing PE leader-follower formation and relaxed bearing rigidity. Section \ref{control} proposes a bearing-based controller and shows that exponential stabilization of the formation is achieved under the bearing PE conditions.
	Section \ref{sec:sim} illustrates the performance of the proposed control strategy on a relaxed rigid formation. The paper concludes with some final comments in Section \ref{conc}.
	\section{Preliminaries} \label{sec:prelmn}
	Let $\mathbb{S}^2:=\{y\in\mathbb{R}^3:\|y\|=1\}$ denote the 2-Sphere and $\|.\|$ the euclidean norm. The operator $[.]_\times$ yields the skew-symmetric matrix associated to its vector argument and  $\lambda_{\max}(.)(\lambda_{\min}(.))$ represents the maximum (minimum) eigenvalue of its matrix argument.
	For any $y\in \mathbb{S}^2$, we can define the projection operator $\pi_y$
	\begin{align*}
	\pi_y := I - y y^{\top} \geq 0, 
	\end{align*}
	which is such that, for any vector $x\in\mathbb{R}^3$, $\pi_y x$ provides the projection of $x$ on the plane orthogonal to $y$. Note that $\pi_y=-[y]_\times[y]_\times$.
	\subsection{Graph Theory}\label{subsec:graph}
	Consider a system of $n\ (n\ge 2)$ connected agents. The underlying interaction topology can be modelled as a digraph (directed graph) $\mathcal{G} := (\mathcal{V}, \mathcal{E})$, where $\mathcal{V}=\{1,2,\ldots,n\}$ is the set of vertices and $\mathcal{E} \subseteq \mathcal{V} \times \mathcal{V}$ is the set of directed edges. In this work, the graph is interpreted as sensing graph, meaning that if the ordered pair $(i,j)\in \mathcal{E}$ then agent $i$ can access or sense information about agent $j$, which is called a neighbor of agent $i$. Note that in a communication graph the information flow would be in the opposite direction. The set of neighbors of agent $i$ is denoted by $\mathcal{N}_i:=\{j\in\mathcal{V}|(i,j)\in\mathcal{E}\}$. Define $m_i=|\mathcal{N}_i|$, where $|.|$ denotes the cardinality of a set.
	A directed path is a finite sequence of distinct vertices $\nu_1,\nu_2,\ldots,\nu_{k-1},\nu_k$, such that $(\nu_{i-1},\nu_i),\ 2\le i \le k$ belongs to $\mathcal{E}$.  A directed cycle is a directed path with the same start and end vertices, i.e. $\nu_1=\nu_k$. A digraph $\mathcal{G}$ is called an acyclic digraph if it has no directed cycle.
	The digraph $\mathcal{G}$ is called a directed tree with a root vertex $i,\ i\in \mathcal{V}$, if for any vertex $j\ne i,\ j\in\mathcal V$, there exists only one directed path connecting $j$ to $i$. Note that a directed tree is acyclic. We say that $\mathcal G$ has a directed spanning tree, if there exists a subgraph of $\mathcal G$ that is a directed tree and contains all the vertices of $\mathcal G$. 
	\subsection {Persistence of Excitation on bearings}
	\begin{definition}
		A positive semi-definite matrix $\Sigma(t)\in \mathbb{R}^{n\times n}$, is called \textit{persistently exciting} (PE) if there exists $T>0$ and $\mu>0$ such that for all $t>0$
		\begin{equation}\scalebox{1}{$
			\frac 1 T\int_{t}^{t+T}\Sigma(\tau)d\tau\ge\mu I. \label{eq:pe}$}
		\end{equation}
		\label{def:pe of matrix}
	\end{definition}
	\vspace{-0.5cm}
	\begin{definition}\label{def:pe}
		A direction $y(t)\in \mathbb{S}^2$ is called \textit{persistently exciting} (PE) if the matrix $\pi_{y(t)}$ satisfies the PE condition according to Definition \ref{def:pe of matrix} with $0<\mu<1$.
	\end{definition}
	%
	\vspace{-0.3cm}
	\begin{lemma}\label{lem:Q_norm}
		Let $Q:=\sum\limits_{i=1}^{l}\pi_{y_i}$. The matrix $Q$ is persistently exciting, if one of the following conditions is satisfied:
		\begin{enumerate}
			\item there is at least one PE direction $y_i$,
			\item there are at least two uniformly non-collinear directions $y_{i}$ and $y_j$, $i,j\in\{1,...,l\},\ i\ne j$. That is:
			$\exists \epsilon_1>0,\; \forall t\ge 0$ such that $|y_i(t)^\top y_j(t)|\le 1-\epsilon_1$.
		\end{enumerate}
	\end{lemma}
	\vspace{-0.6cm}
	\begin{pf}
		The proof is given in \cite[Lemma 3]{le2017observers}.
	\end{pf}
	\section{Bearing PE leader-follower formation and relaxed bearing rigidity}\label{sec:persistent}
\begin{figure}[!t]
		\centering
		\includegraphics[width=1.2in]{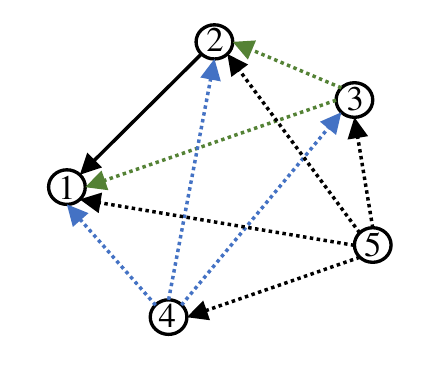}	
		\caption{Possible connections of a leader-follower structure when $n=5$. The solid line represents the unique neighbor of the first follower (agent 2) which is the leader (agent 1). The dashed lines represent all possible connections of the followers 2, 3, and 4.}
		\label{fig:path}
	\end{figure}

	\begin{definition} \label{def:relaxedLFF}
		A digraph $\mathcal{G}=(\mathcal V,\mathcal E)$ has a leader-follower structure if it is acyclic and has a directed spanning tree. It has a minimal leader-follower structure if each follower $i$ ($i\in \mathcal V,\ i\ne 1$) has only one neighbor.
	\end{definition}
	The leader-follower structure defined above is more general than the leader-first follower structure (LFF) considered in \cite{trinh2019bearing}, for which each follower has two neighbors except the first follower which is only connected to the leader.
	In our setting, the leader is the root vertex which has no neighbors and each of the other followers has at least one neighbor. Without loss of generality, the agents are numbered (or can be renumbered) such that agent $1$ is the leader, i.e.  $\mathcal{N}_1= \varnothing$, agent $2$ is the first follower with $\mathcal{N}_2 = \{1\}$, and for each agent $i\geq3$ the set of neighbors satisfies $\mathcal{N}_i \subseteq\{1,\ldots,i-1\}$. An example of a possible 5-agent leader-follower graph is shown in the Figure \ref{fig:path}. 
	
	%
	Given a digraph $\mathcal{G}$, 
	let $p_i  \in \mathbb{R}^3$ denote the position and $v_i  \in \mathbb{R}^3$ the velocity of each agent $i\in \mathcal V$, both expressed in an inertial frame common to all agents, such that $\dot p_i=v_i$. The stacked vector $\boldsymbol{p}=[p_1^\top,...,p_n^\top]^\top\in \mathbb{R}^{3n}$ is called a configuration of $\mathcal{G}$ and the digraph $\mathcal{G}$ together with the configuration $\boldsymbol{p}$ define a formation $\mathcal{G}(p)$ in the 3-dimensional space.
	Defining the relative position vectors
	\begin{equation}
	\label{eq:eij}
	p_{ij}:=p_{j}-p_i, \ i,\ j \in \mathcal{V},\ i\ne j
	\end{equation}
	and as long as $\|p_{ij}\|\neq 0$, the bearing of agent $j$ relative to agent $i$ is given by the unit vector
	\begin{equation}
	\label{eq:gij}
	g_{ij} := p_{ij}/\|p_{ij}\| \in \mathbb{S}^2.
	\end{equation}
	Similarly to $p_{ij}$, define $v_{ij}:=v_j-v_i$ as the relative velocity between agent $i$ and $j$.
	
	\begin{definition}\label{def:pe_formation}
		A formation $\mathcal{G}(p(t))$ is called bearing persistently exciting, if $\forall i\in V$, the matrices $\sum\limits_{j\in\mathcal{N}_i}\pi_{g_{ij}(t)}$ satisfy the PE condition.
	\end{definition}
	The following Theorem shows that a leader-follower formation can be uniquely determined if it is bearing PE.
	\begin{thm}\label{thm:shape}
		Consider a leader-follower formation. Assume that the leader's position $p_1(t)$, its velocity $v_1(t)$, the bearing vectors $\{g_{ij}(t)\}_{(i,j)\in \mathcal{E}}$, and the corresponding relative velocity vectors $\{v_{ij}(t)\}_{(i,j)\in \mathcal{E}}$ (equivalently $v_i(t)$) are well-defined, known, and bounded.  Let $\hat p_1 \triangleq p_1$ and ${\hat p}_i$ denote the estimate of $ p_i$, for $i=2,\ldots, n$ with the following dynamics:
		\begin{equation}\scalebox{1}{$  \label{eq:observer}
			\dot{\hat p}_i=v_i-K\sum\limits_{j\in\mathcal{N}_i}\pi_{g_{ij}}(\hat p_i-\hat p_j),\ \forall i\ge 2,$}
		\end{equation}
		with arbitrary initial conditions and $K$ a positive definite matrix. Assume that the leader-follower formation is bearing persistently exciting. Then $\hat{p}_i$ converges uniformly globally exponentially (UGE) to the unique $p_i$.
	\end{thm}
	\vspace{-0.6cm}
	\begin{pf}
		Consider the error variables $\tilde p_i := \hat p_i - p_i$ defined for $i=2,\ldots, n$ and the corresponding dynamics obtained from \eqref{eq:observer}. For $i=2$, we have $\mathcal{N}_2=\{1\}$ and it is straightforward to verify that the dynamics of $\tilde p_2$ is given by
		\begin{equation}\scalebox{1}{$ \label{eq:dot_p2}
			\dot {\tilde p}_2 = - K \pi_{g_{21}}\tilde p_2$}
		\end{equation}
		and that $\tilde p_2=0$ is UGE stable under the PE condition (by direct application of \cite[Lemma 4]{le2017observers}).
		For $i=3$ and $\mathcal{N}_3=\{1\}$, the proof is exactly the same as for agent 2. For $\mathcal{N}_3=\{2\}$ or $\{1,2\}$, the dynamics of ${\tilde p}_3$ can be written as
		\begin{equation} \scalebox{1}{$ \label{eq:dot_p3}
			\dot {\tilde p}_3 = - K \sum\limits_{j\in\mathcal{N}_3} \pi_{g_{3j}} \tilde p_3 + K \pi_{g_{32}} \tilde p_2 $}
		\end{equation}
		which together with \eqref{eq:dot_p2} forms a cascaded system with $\tilde p_2$ as input to \eqref{eq:dot_p3}.
		Using the fact that $\tilde p_2=0$ is UGE stable and system \eqref{eq:dot_p3} is continuously differentiable and globally Lipschitz in $(\tilde p_3, \tilde p_2)$, it follows (by direct application of \cite[Proposition 1]{le2017observers}) that $\tilde p_3=0$ is also UGE stable.
		In the general case, we can write
		\begin{equation}\scalebox{1}{$  \label{eq:dot_pi}
			\dot {\tilde p}_i = - K \sum\limits_{j\in\mathcal{N}_i} \pi_{g_{ij}} \tilde p_i
			+ K \sum\limits_{j\in\mathcal{N}_i\setminus\{1\}}\pi_{g_{ij}} \tilde p_j, $}
		\end{equation}
		for $i=2,\ldots, n$ and the proof of that $\tilde p_i = 0$ is UGE stable can be obtained in a similar way.

	\end{pf}
	
	\begin{remark}
		For the static case where ${v}_{ij}=0,\ \forall (i,j)\in \mathcal E$, we obviously conclude that $g_{21}$ is not PE. In that case, if each agent $i\ (i\ge 3)$ has two neighbors $1\le j\ne k< i$ with $g_{ij}\ne \pm g_{ik}$,  the leader-follower formation becomes exactly the same as the bearing rigid desired LFF  formation described in \cite{trinh2019bearing} and uniqueness of the formation can still be guaranteed if, for instance, the distance $d_{21}=\|p_1-p_2\|$ is provided. Under the proposed controller, which will be defined in the next section, the formation will converge to the desired shape up to a scaling factor as discussed in \cite{trinh2019bearing}.
		
		Note that under the condition of Theorem \ref{thm:shape}, the shape and the size of the bearing PE leader-follower formation may be time-varying. This includes similarity transformations (a combination of rigid transformation and scaling) involving a time-varying rotation. In this case, it is straightforward to show that for any bearing formation the bearing measurements are invariant to translation and scaling but change with rotation such that $g_{ij}(t)=R(t)^\top g_{ij}(0),\forall (i,j)\in\mathcal{E}$ (with $R(t) \in SO(3)$ the rotation part of the similarity transformation). This implies that there exists similarity transformations in which $R(t)$ is time-varying such that the leader-follower formation $\mathcal G(p(t))$ is bearing PE.
		
	\end{remark}	
	\begin{definition}
		A leader-follower formation $\mathcal G(p(t))$ is called relaxed bearing rigid if it is bearing PE and subjected to a similarity transformation.
	\end{definition}
	\begin{corollary}
		If the formation is relaxed bearing rigid, then the result of Theorem \ref{thm:shape} applies.
	\end{corollary}
	\vspace{-0.65cm}
	\begin{pf} 
		The proof is analogous to the proof of Theorem \ref{thm:shape}. It is omitted here for the sake of brevity.
	\end{pf}
	\section{Bearing formation control}\label{control}
	Consider the formation $\mathcal{G}(p)$, where each agent $i\in\mathcal{V}$ is more realistically modeled as double integrator with the following dynamics:
	\begin{equation}\label{eq:double integrator}
	\left\{
	\begin{aligned}
	\dot{p}_i&=v_i\\
	\dot{v}_i&=u_i
	\end{aligned}
	\right.
	\end{equation}
where $u_i\in\mathbb{R}^3$ is the acceleration control input expressed in the inertial frame. Let $p_i^*(t)$, $v_i^*(t)$ and $u^*_i(t) \in\mathbb{R}^3$ denote the desired position, velocity, and acceleration of agent $i$, respectively, and define the desired relative position vectors $p_{ij}^*$ and bearings $g_{ij}^*$, according to \eqref{eq:eij} and \eqref{eq:gij}, respectively.
	
	We assume that the $n$-agent system satisfies the following assumptions.
	\begin{assumption}\label{ass:desired} The desired acceleration $u_i^*(t)$ and the desired relative velocity $v_{ij}^*(t)$ are bounded for all $t>0$, the resulting desired bearings $g_{ij}^*(t)$ are well-defined for all $t>0$  and the desired formation is bearing PE.
	\end{assumption}
	\begin{assumption} \label{ass:construction}
		The sensing topology of the group is described by a digraph $\mathcal{G}(\mathcal{V},\mathcal{E})$ that satisfies the leader-follower structure defined in Definition \ref{def:relaxedLFF}. Each agent $i\ge 2$ can measure the relative velocity $v_{ij}$ and relative bearing vectors $g_{ij}$ to its neighbors $ j\in \mathcal{N}_i$.
	\end{assumption}

	
	\begin{assumption}\label{ass:collision}
		As the formation evolves in time, no inter-agent collisions and occlusions occur. In particular, we assume that the bearing information $ g_{ij}(t),\ (i,j)\in\mathcal{E}$ is all the time well-defined.
	\end{assumption}
	
	With all these ingredients, we can define the bearing formation control problem as follows.
	\begin{problem}
		Consider the system \eqref{eq:double integrator} and the formation $\mathcal{G}(p)$. Under Assumptions \ref{ass:desired}-\ref{ass:collision}, design stabilizing distributed control laws based on bearing and relative velocity measurements that guarantee exponential stabilization of the formation in terms of shape and scale  to the desired one.
	\end{problem}
	For any agent $i \ (i\ge 2)$ and any agent $j$ in its neighbor ($j \in\mathcal N_i$), we define the relative position error $\tilde{p}_{ij}:=p_{ij}-p_{ij}^*$ and the relative velocity error $\tilde{v}_{ij}:=\dot{\tilde{p}}_{ij}=(v_j-v_i)-(v_j^*-v_i^*)$ along with the following dynamics:
	\begin{equation}\scalebox{1}{$ \label{eq:states_f}
		\left\{
		\begin{aligned}
		\dot{\tilde{p}}_{ij}&=\tilde{v}_{ij}\\
		\dot{\tilde{v}}_{ij}&=u_j-u_j^*-(u_i-u_i^*).\\
		\end{aligned}
		\right.$}
	\end{equation}
	Consider the following control law for each agent $i\in\mathcal{V}$
	\begin{equation}\scalebox{1}{$
		u_i=\sum\limits_{j\in\mathcal{N}_i}[-k_{p_i}\pi_{g_{ij}}p_{ij}^*+k_{d_i}\tilde{v}_{ij}]+u_i^*, \label{eq:ui}$}
	\end{equation}
	where $k_{d_i}$ and $k_{p_i}$ are positive gains that satisfy $k_{d_i}> \frac{1}{m_i}$ and $k_{p_i}<\frac 4 {m_i}-\frac{4}{k_{d_i}^2m_i^3}$ (recall that $m_i = |\mathcal{N}_i|$). For $i\in \mathcal{V}\backslash\{1\}$, define new variables $ \tilde{x}_{ij}:=(\tilde{p}_{ij}^\top,\tilde{v}_{ij}^\top)^\top,\ j\in\mathcal N_i$ and the following matrices to be used later in the stability analysis:
	\begin{equation} \label{eq:Ai_Pi}\scalebox{0.85}{$
		A_{i}(g_i)=\begin{bmatrix}
		0     & -I \\
		k_{p_i}\sum\limits_{l\in\mathcal{N}_i}\pi_{g_{il}}& {k_{d_i}m_i}I\\
		\end{bmatrix},\  P_{i}:=\frac12 \begin{bmatrix} I & \frac 1 {{k_{d_i}m_i}}I\\\frac 1 {{k_{d_i}m_i}}I& I \end{bmatrix} >0, $}
	\end{equation}
	\begin{equation} \label{eq:Qi_Sigmai}\scalebox{0.75}{$
		Q_{i}(g_i)=\sum\limits_{j\in\mathcal{N}_i}\begin{bmatrix}
		\frac{k_{p_i}}{{k_{d_i}m_i}}\pi_{g_{ij}} &\frac {k_{p_i}} 2 \pi_{g_{ij}}\\\frac {k_{p_i}} 2 \pi_{g_{ij}}&({k_{d_i}}-\frac{1}{{k_{d_i}m_i^2}})I
		\end{bmatrix} \geq 0,\ \mbox{and}\  \Sigma_i=\begin{bmatrix}
		\sum\limits_{j\in\mathcal{N}_i}\pi_{g_{ij}^*}& 0\\0& I
		\end{bmatrix}\ge 0,$}
	\end{equation}
	where the matrices argument $g_i$ stands for the concatenation of all bearing vectors $g_{ij},\ \forall j\in \mathcal N_i$.
	\subsection{Stability and convergence of the first follower}
	
	\begin{lemma} \label{lem:2agent}
		Consider a $n$-agent ($n\ge 2$) system with a leader-follower interaction topology as specified in Definition \ref{def:relaxedLFF}. For the first follower ($i=2$), consider the error dynamics \eqref{eq:states_f} along with the control law \eqref{eq:ui}.
		If Assumptions \ref{ass:desired}-\ref{ass:collision} are satisfied, then the equilibrium point $\tilde  x_{21}=(\tilde{p}_{21}^\top,\tilde{v}_{21}^\top)^\top=0$ is  exponentially stable (ES).
	\end{lemma}
	\vspace{-0.5cm}
	\begin{pf}
		Recalling \eqref{eq:states_f} and \eqref{eq:ui}, the closed-loop system for the state $\tilde x_{21}$ is expressed as
		\begin{equation}\scalebox{1}{$
			\dot {\tilde x}_{21}=-A_{2}(g_{2}(t))\tilde x_{21}. \label{eq:dotx_1}$}
		\end{equation}
		Consider the following Lyapunov function candidate:\\
		\begin{equation}\scalebox{1}{$
			\begin{aligned}
			\mathcal{L}_{21}=\tilde x_{21}^\top P_{2} \tilde x_{21}.
			\end{aligned}$}
		\end{equation}
		Taking its time-derivative yields
		\begin{equation}\scalebox{1}{$ \label{eq:dotL1}
			\begin{aligned}
			\dot{\mathcal{L}}_{21}=-\tilde x_{21}^\top Q_{2}\tilde x_{21}.
			\end{aligned}$}
		\end{equation}
		Since $Q_{2}$ is positive-semidefinite, one concludes that the state $\tilde x_{21}$ is bounded.
		By direct application of Lemma \ref{lem:Matirx PE} (see appendix), it is straightforward to verify that
		\begin{equation}\scalebox{1}{$
			\begin{aligned}
			\dot{\mathcal{L}_{21}}
			=-\tilde x_{21}^\top Q_{2}\tilde x_{21}\leq-\gamma_2 \tilde x_{21}^\top \Sigma_2 \tilde x_{21}\leq 0,
			\end{aligned} \label{eq:dotL1le}$}
		\end{equation}
		where $\gamma_2$ can be deduced from the proof of Lemma \ref{lem:Matirx PE}:
		\begin{equation*}
		\scalebox{1}{$\gamma_2=\frac{k_{p_2}(k_{d_2}-k_{p_2}k_{d_2}^2/4-1)}{k_{d_2}(k_{p_2}+k_{d_2}-1)}\min\alpha_2^2(t)>0  $}\end{equation*}
		with $ \min\alpha_2^2(t)=\min\frac{\|p_{21}^*(t)\|^2}{\|p_{21}(t)\|^2}$. Now, since $\mathcal{L}_{21}$ is decreasing, one can verify that
		\begin{equation*}\scalebox{1}{$
			\min\alpha_2^2(t)\geq\frac{\min\|p_{21}^*(t)\|^2}{(\sqrt{\frac{\lambda_{\max}(P_2)}{\lambda_{\min}(P_2)}}\|\tilde x_{21}(0)\|+\max\|p_{21}^*(t)\|)^2}.$}
		\end{equation*}
		From \eqref{eq:Qi_Sigmai} along with the PE condition of $g_{21}^*$, one ensures that condition (1) of Theorem \ref{thm:ES} in the appendix is satisfied. By a direct application of Lemma \ref{lem:c} (see appendix) one can conclude that condition (2) of Theorem \ref{thm:ES} is also satisfied. This in turn implies that
		$\tilde x_{21}=0$ is ES.
	\end{pf}
	\begin{remark}
		Note that in the above lemma, Assumption \ref{ass:collision} relies on the evolution of state variables. This assumption serves here to show that if there is no collision or occlusion, the bearings are well-defined and the proposed control design yields the desired convergence properties (Lemma \ref{lem:2agent} and even in the following results: Lemma \ref{lem:3agent} and Theorem \ref{thm:fullstate}). Trying to more specifically characterize the set of initial conditions for which the system's solutions avoid collision and occlusion is out of the scope of the paper.
	\end{remark}
	\subsection{Stability and convergence of the second follower}
	\begin{lemma} \label{lem:3agent}
		Consider a $n$-agent ($n\ge 3$) system with a leader-follower interaction topology as specified in Definition \ref{def:relaxedLFF}. For the second follower ($i=3$), consider the error dynamics \eqref{eq:states_f} along with the control law \eqref{eq:ui}.
		If the Assumptions \ref{ass:desired}-\ref{ass:collision} are satisfied and Lemma \ref{lem:2agent} is valid, then the equilibrium point $\tilde x_{3j}=(\tilde{p}_{3j}^\top,\tilde{v}_{3j}^\top)^\top=0,\ \forall j\in \mathcal{N}_3$ is ES.\\
	\end{lemma}
	\vspace{-1.1cm}
	\begin{pf}	
		According to the leader-follower structure described in Definition \ref{def:relaxedLFF}, the second follower (agent $3$) can have three possible sets of neighbors: $\mathcal{N}_3=\{1\}$,  $\mathcal{N}_3=\{2\}$ and $\mathcal{N}_3=\{1,2\}$.
		
		Case i): $\mathcal{N}_3=\{1\}$, the proof is identical to the proof of Lemma \ref{lem:2agent}.
		
		Case ii):
		$\mathcal{N}_3=\{2\}$ or $\mathcal{N}_3=\{1,2\}$.
		Since $\tilde x_{31}=\tilde x_{32}+\tilde x_{21}$,
		the closed-loop system for the states $\tilde x_{3j},\ j\in\mathcal{N}_3$ is expressed as
		\begin{equation}\scalebox{1}{$  \label{eq:3states_12}
			\dot {\tilde x}_{3j}=-A_{3}(g_{3}(t))\tilde x_{3j}+B_{21}(g_{3}(t),g_{2}(t))\tilde x_{21}$}
		\end{equation}
		where  $A_3$ is defined in \eqref{eq:Ai_Pi} and $B_{21}$ is a bounded function.
		We can interpret \eqref{eq:3states_12} as a cascaded system that has $\tilde x_{21}$ as input to the unforced system
		\begin{equation}\scalebox{1}{$  \label{eq:3states_12_unforce}
			\dot {\tilde x}_{3j}=-A_{3}(g_{3}(t))\tilde x_{3j}.$}
		\end{equation}
		Now the proof becomes analogous to the proof of Lemma \ref{lem:2agent}.  By a direct application of Theorem 3, one concludes that the equilibrium $\tilde x_{3j}=0,\ j\in \mathcal{N}_3$ of the unforced system \eqref{eq:3states_12_unforce} is ES. Since the matrix valued function $B_{21}$ is bounded and $\tilde x_{21}=0$ is ES, this implies that the equilibrium point $\tilde x_{3j}=0,\ j\in \mathcal{N}_3$ is ES for the system \eqref{eq:3states_12}.
	\end{pf}
	\vspace{-0.3cm}
	\subsection{The $n$-agent system}
	\begin{thm}\label{thm:fullstate}
		Consider a $n$-agent ($n\ge 2$) system with a leader-follower interaction topology as specified in Definition \ref{def:relaxedLFF}. For all agents $i\in\mathcal{V}\backslash\{1\}$, consider the system \eqref{eq:states_f} along with the proposed control law \eqref{eq:ui}.
		If Assumptions \ref{ass:desired}-\ref{ass:collision} are satisfied, then the equilibrium point $\tilde{x}_{ij}=(\tilde{p}_{ij}^\top,\tilde{v}_{ij}^\top)^\top=0$ is ES, $\ \forall i\in \mathcal{V}\backslash\{1\} $ and $\forall j\in \mathcal{N}_i$.
	\end{thm}
	\vspace{-0.6cm}
	\begin{pf}
		We will prove the convergence of $\tilde{x}_{ij}=0$ by mathematical induction. Firstly, for $i=2$ and $i=3$ the conclusion that $\tilde x_{ij}=0$ is ES $\forall j\in \mathcal{N}_i$ follows directly from Lemma  \ref{lem:2agent} and Lemma \ref{lem:3agent}, respectively.
		Secondly, we suppose that $\tilde x_{kj}=0$ is ES,  $\forall j\in \mathcal{N}_k$ and $\forall 4\le k \le i-1$ then, we show that it is also true for $k=i$. Using the fact that $\forall q\in\mathcal N_i$, one has $ \tilde{x}_{iq}=\tilde x_{ij}+\tilde x_{jq}$  with $j\in \mathcal N_i, j\ne q$ and $\tilde x_{jq}$ can be expressed in terms of the error variables $\tilde x_{km},\ 2\le k \le i-1,\ m\in\mathcal{N}_k$ because  the graph is connected, the closed-loop system  for the states $\tilde{x}_{ij},\ j\in \mathcal{N}_i$ can be represented as
		\begin{equation} \label{eq:states_i_re}\scalebox{0.85}{$
			\begin{aligned}
			\dot {\tilde{x}}_{ij}=& -A_i(g_{i}(t))\tilde{x}_{ij}+\sum\limits_{2\le k\le i-1,\ m\in \mathcal N_k} B_{km}(g_{i}(t),g_{k}(t))\tilde x_{km}
			\end{aligned}$}
		\end{equation}
		where $A_i$ is defined in \eqref{eq:Ai_Pi} and $B_{km}$ is a bounded matrix valued function.
		Thus system \eqref{eq:states_i_re} can be considered as a cascaded system with $\tilde x_{km},\ 2\le k\le i-1,\ m\in \mathcal N_k$ perturbing the unforced system $\dot {\tilde{x}}_{ij}=-A_i(g_{i}(t)) \tilde{x}_{ij}$.
		From there and analogously to Lemma \ref{lem:2agent} and \ref{lem:3agent}, one concludes that  $\tilde x_{ij} = 0$ is ES for the unforced system. Because the error variables $\tilde x_{km}=0$ are ES and $B_{km}$ is bounded, $\tilde x_{ij}=0$ is also ES for system (22).
		Then, by mathematical induction, it follows that the claim is true for all $i\in\mathcal{V}\backslash\{1\}$, which concludes the proof.
	\end{pf}
	It is worth to notice that the exponential stabilization of the equilibrium $(\tilde{p}_{ij}^\top,\tilde{v}_{ij}^\top)=0, \ \forall(i,j)\in \mathcal E$ implies the exponential stabilization of the formation to the desired one in terms of shape and scale only. This is inherent to the problem at hand since only relative measurements are involved in the control design. However, by exploiting the cascade structure of the formation dynamics, it is straightforward to verify that the exponential stabilization of the formation in the configuration space  (that is $p_i \rightarrow p_i^*$) can be directly deduced if the leader has access to its own position.
	\section{Simulation Results}\label{sec:sim}
\begin{figure}[!htb]
		\centering
		\includegraphics[scale = 0.6]{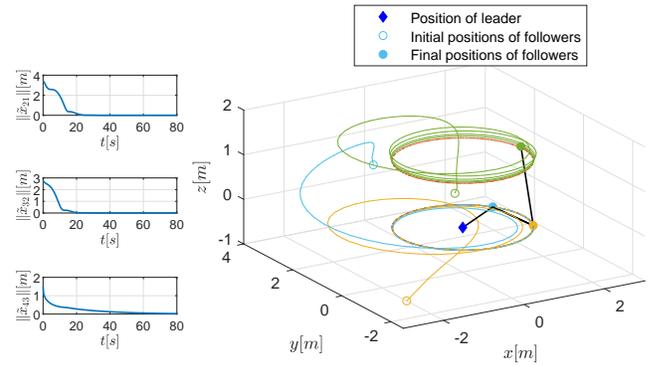}	
		\caption{Evolutions of error states (left hand side) and 3-D trajectories (right hand side) for a pyramid formation under a minimal leader-follower structure: the colored solid lines represent the agents' trajectories, the  dashed red lines represent the desired trajectories and the black solid lines represent the connections between agents.}
		\label{fig:3D2}
	\end{figure}

	In this section, we consider a four-agent system defined in $\mathbb{R}^3$, $\mathcal{V} = \{1,2,3,4\}$, with a minimal leader-follower graph formed by a single directed path, that is, each follower has only one neighbor such that $\mathcal{N}_i=\{i-1\},\ i\in\mathcal{V}\backslash\{1\}$.
	For the sake of simplicity,
	the leader (agent 1) is static at position $p_1=[0\ 0\ 0]^\top$. According to Assumption \ref{ass:desired}, the desired trajectories for the followers are chosen such that $p_i^* (t) = R(t)^\top p_i^*(0)$, with \scalebox{0.7}{$R(t)=\begin{bmatrix}
			\cos(\frac t {2.5}) & -\sin(\frac t {2.5}) &0\\\sin(\frac t {2.5}) &\cos(\frac t {2.5})& 0\\ 0 &0& 1
			\end{bmatrix}$}, $p_2^*(0)=[0\ 1 \ 0]^\top, p_3^*(0)=[\frac {\sqrt 3} 2\ \frac 1 2 \ 0]^\top$ and $p_4^*(0)=[\frac 1 2 \ \frac {\sqrt 3} 2\  1]^\top$, which form a pyramid in $\mathbb{R}^3$ that rotates about $z$-axis (see Fig.~\ref{fig:3D2}). Note that the desired formation is not bearing rigid but relaxed bearing rigid. The initial conditions are $p_2(0)=[-1\ 2 \ 1]^\top$, $v_2(0)=[0\ 1 \ 0]^\top$, $p_3(0)=[-2\ -1 \ -1]^\top,\ v_3(0)=[1\ 0 \ 0]^\top, p_4(0)=[-0.5\ -0.5 \ 1]^\top$ and $v_4(0)=[1\ 0 \ -1]^\top$. The controller gains are chosen as follows $ k_{p_i}=3$ and $k_{d_i}=10, \forall i\in \mathcal V\backslash\{1\}$, to ensure a fast convergence rate according to Theorem \ref{thm:ES} while ensuring that inequalities $k_{d_i}> 1$ and $k_{p_i}<4-\frac{4}{k_{d_i}^2}$ are satisfied.
	The left hand side of Fig. \ref{fig:3D2} shows the time evolution of the error states $\|\tilde{x}_{21}\|$, $\|\tilde{x}_{32}\|$ and $\|\tilde{x}_{43}\|$, respectively. It also confirms the result of Proposition \ref{prop:rate} that due to the cascade structure of the system the convergence of $\tilde{x}_{21}(t)$ is the fastest and of $\tilde{x}_{43}(t)$ is the slowest one. The right hand side of Fig. \ref{fig:3D2} shows the 3-D time evolution of the formation converging to the desired one. It also validates the fact that the proposed control law stabilizes the formation without requiring bearing rigidity (additional simulation results with animations can be found in \url{https://youtu.be/fwv4Q_3xCWw}). 
	\vspace{-0.3cm}
\vspace{-0.4cm}
	\section{Conclusion}\label{conc}
This paper studies bearing formation control problem of a leader-follower structure under time-varying desired formation and introduces the new concept of relaxed bearing rigidity.  The proposed controller ensures, a (local) exponential stability of the formation as long as the bearing PE conditions are met on the desired formation. Simulation results are provided to validate the control strategy. Future work will be dedicated to the incorporation of collision avoidance to bypass Assumption \ref{ass:collision} and to ensure at least semi-global exponential stability.
	
	\begin{ack}                               
		 This work was partially supported by the Project MYRG2018-00198-FST of the University
		 of Macau; by the Macao Science and Technology, Development Fund under Grant FDCT/0031/2020/AFJ; by Funda\c{c}\~{a}o  para a Ci\^{e}ncia e a Tecnologia (FCT) through Project UIDB/50009/2020 and PTDC/EEI-AUT/31411/2017;
		 and by the ANR-DACAR project. The work of Z. Tang was supported by FCT through Ph.D. Fellowship PD/BD/114431/2016 under the FCT-IST NetSys Doctoral Program.  
	\end{ack}

\bibliographystyle{agsm}

\bibliography{bibliography}
	
	\appendix
	\section{Technical Lemmas and Theorem }    
	\begin{lemma} \label{lem:Matirx PE}
		Let $x_1(t),  x_1^*(t)\in \mathbb{R}^3/\{0\}$ and $x_2(t),x_2^*(t) \in \mathbb{R}^3$ be bounded functions of time and $x_1(t)$ and $x_1^*(t)$ are such that $y_1=\frac {x_1} {\|x_1\|},\ y_1^*=\frac{x_1^*}{\|x_1^*\|}$ are well defined. Define $\tilde{x}=[(x_1-x_1^*)^\top,(x_2-x_2^*)^\top]^\top\in\mathbb{R}^6, $ and the matrix-valued functions
		\vspace{-0.3cm}	
		\begin{equation*}\scalebox{0.85}{$
			\Gamma(t)=\begin{bmatrix}
			c_3\pi_{y_1} & c_2 \pi_{y_1} \\
			c_2\pi_{y_1} & c_1I
			\end{bmatrix}$}
		\end{equation*}
		where $c_1$, $c_2$, and $c_3$ are positive constants. If $c_1c_3> c_2^2>\epsilon_1$ with $\epsilon_1>0$, then $\tilde{x}^\top\Gamma(t)\tilde{x}\ge\gamma\tilde{x}^\top\Sigma(t)\tilde{x}$, with \scalebox{0.8}{$\Sigma(t)=\begin{bmatrix} \pi_{y_1^*(t)} & 0 \\ 0 & I\end{bmatrix}$} and $\gamma$ a positive constant.
	\end{lemma}
	\vspace{-0.7cm}
	\begin{pf}
		Since $y_1^{*\top}\pi_{y_1}y_1^*=y_1^\top\pi_{y_1^*}y_1$, it is straightforward to verify that $\tilde{x}\Gamma\tilde{x}=\tilde{x}\Gamma^{'}\tilde{x}$ with
		\begin{equation*}\scalebox{0.85}{$
			\Gamma^{'}(t)=\begin{bmatrix}
			\alpha^2c_3\pi_{y_1^*} & -\alpha_i c_2 [y_1^*]_\times[y_1]_\times \\
			-\alpha c_2[y_1]_\times[y_1^*]_\times & c_1I
			\end{bmatrix},\ \alpha=\frac{\|x_1^*\|}{\|x_1\|}.$}
		\end{equation*}
		Note that $\Gamma^{'} = S^\top \Lambda M \Lambda S$, where \scalebox{0.8}{$S = \begin{bmatrix} [y_1^*]_\times & 0 \\ 0 & I\end{bmatrix}$, $\Lambda=\begin{bmatrix}
			\alpha I &0 \\0& I
			\end{bmatrix}$} and \scalebox{0.8}{$M=\begin{bmatrix} c_3 I& c_2[y_1]_{\times}\\-c_2[y_1]_{\times}& c_1 I \end{bmatrix}$}. Due to the fact that $c_1c_3> c_2^2>\epsilon_1$, $M\ge \lambda_M I>0$ with $\lambda_M=\frac{c_3c_1-c_2^2}{c_3+c_1}$. Thus one has $\Gamma^{'}\ge \lambda_MS^\top \Lambda^2 S$.
		Now using the fact that $\alpha(t)=\frac{\|x_1^*(t)\|}{\|x_1(t)\|} \ge \frac{\min \|x_1^*(t)\|}{\max\{\|x_1(t)-x_1^*(t)\|+\|x_1^*(t)\|\}}$, we can conclude that $\tilde{x}^\top\Gamma\tilde{x}=\tilde{x}^\top\Gamma^{'}\tilde{x}\ge \gamma\tilde{x}^\top\Sigma(t)\tilde{x} $ with \scalebox{0.8}{$\Sigma(t)=\begin{bmatrix} \pi_{y_1^*(t)} & 0 \\ 0 & I\end{bmatrix}$} and \scalebox{0.9}{$\gamma=\lambda_M\min\alpha^2(t)\ge \lambda_M\frac{\min \|x_1^*(t)\|^2}{\max(\|x_1(t)-x_1^*(t)\|+\|x_1^*(t)\|)^2}>0$}.
		
	\end{pf}

	\begin{lemma} \label{lem:c}
		Let $y_i\in \mathbb{S}^2,\ i=1,\ldots,m$ and define the matrix-valued functions
		\begin{equation*}\scalebox{0.85}{$
			A=\begin{bmatrix}
			0     & -I \\
			c_{4}\sum\limits_{i=1}^m\pi_{y_i}& c_{5}I\\
			\end{bmatrix} \mbox { and } Q=\begin{bmatrix}
			c_3\sum\limits_{i=1}^m\pi_{y_i} &c_2 \sum\limits_{i=1}^m\pi_{y_i}\\c_2 \sum\limits_{i=1}^m\pi_{y_i}&mc_1I
			\end{bmatrix}$}
		\end{equation*}
		where $c_1$, $c_2$, $c_3$, $c_4$ and $c_5$ are positive constants, such that $c_3c_5>c_4^2$. There exists $c>0$ such that $cQ-A^\top A\geq 0$.
	\end{lemma}
	\vspace{-0.8cm}
	\begin{pf}
		Define \scalebox{0.85}{$H_i=\begin{bmatrix}\pi_{y_i} & 0\\0 & I \end{bmatrix},\ H =\sum\limits_{i=1}^m H_i= \begin{bmatrix} \sum\limits_{i=1}^m\pi_{y_i} & 0\\0 & mI \end{bmatrix},$}
		\scalebox{0.85}{$l_Q = \lambda_{\min} (\begin{bmatrix}
			c_3 &c_2 \\c_2 &c_1
			\end{bmatrix}),\text{and}\ l_A = \lambda_{\max} (\begin{bmatrix}
			c_4^2 &c_5 c_4 \\c_5c_4 & c_5^2+1
			\end{bmatrix})$}. Since\\ \scalebox{0.85}{$Q=\sum\limits_{i=1}^{m}H_i\begin{bmatrix}
			c_3 &c_2 \\c_2 &c_1
			\end{bmatrix}H_i$} \scalebox{0.85}{$\ge l_Q\sum\limits_{i=1}^{m}H_i^2=l_Q\sum\limits_{i=1}^{m}H_i= l_Q H$}
		and \\ \scalebox{0.85}{$ A^{\top} A =H\begin{bmatrix}
			c_4^2 &c_5 c_4 \\c_5c_4 & c_5^2+1
			\end{bmatrix} H\leq l_A H^2
			$}, we can conclude that $cQ-A^\top A\geq 0$ if $c l_Q - l_A \lambda_{\max}(H)\geq0$, which holds if $c \geq \frac{l_A}{l_Q}m$.
	\end{pf}
	\begin{thm}\label{thm:ES}
		Consider the following system
		\begin{equation}\label{dot x}\scalebox{0.9}{$
			\dot x(t)=f(x(t),t),\ x \in \mathbb{R}^n$}
		\end{equation}
		with $f(x(t),t)$ a piecewise continuous and locally Lipschitz function such that $f(0,t)=0$. Assume there exists a function $\mathcal L_x(t)=\mathcal L (t,x(t))\in \mathbb R^+$, such that $\lambda_1\|x(t)\|^2\le \mathcal{L}_x(t)\le \lambda_2\|x(t)\|^2$ and $\dot{\mathcal{L}}_x(t)\le -\gamma x(t)^\top \Sigma(t) x(t)$, where $\Sigma(t)\in \mathbb{R}^{n\times n}$ is an upper bounded  positive semi-definite function $(\|\Sigma(t)\|\le \lambda_{\Sigma})$, with $\lambda_1$, $\lambda_2$, $\lambda_{\Sigma}$ positive constants and $\gamma(x(0))$ a positive function of the initial state $x(0)$. If\\
		1) $\Sigma(t)$ satisfies the PE condition in Definition \ref{def:pe of matrix} and,\\
		2) $\dot {\mathcal{L}}_x(t)\le -\frac 1 c \|f(x,t)\|^2\le 0$, $c>0$,\\
		then the origin of \eqref{dot x} is exponentially stable (ES), and verifies: $x(t)\le \sqrt{\frac{\lambda_2}{\lambda_1(1-\sigma)}}x(0)\exp(-\frac \sigma {2T} t)$ with $\sigma = \frac{1}{1+\rho}\frac{1}{1+ \rho c T^2\gamma\lambda_\Sigma}$ and $\rho = \frac{\lambda_2}{\mu T\gamma}$.
	\end{thm}
	\vspace{-0.6cm}
	\begin{pf}
		The proof follows the arguments used in \cite[Lemma 5]{loria2002uniform}. Taking integral of  $\dot{\mathcal{L}}_x(t)\le -\gamma x(t)^\top \Sigma(t) x(t)$, we get
		\begin{equation}\scalebox{0.9}{$ \label{eq:intL21}
			\mathcal{L}_x(t+T)-\mathcal{L}_x(t)\le-\gamma\int_{t}^{t+T}\|\Sigma^{\frac1 2}(\tau){x}(\tau)\|^2 d \tau$}
		\end{equation}
		where, according to \eqref{dot x}, $x(\tau)$ can be rewritten as
		\begin{equation}\scalebox{0.9}{$\label{int_x21}
			x(\tau)=x(t)+\int_{t}^{\tau}f(x(s),s)ds.$}
		\end{equation}
		Substituting \eqref{int_x21} in \eqref{eq:intL21} and using $\|a+b\|^2\ge[\rho/(1+\rho)]\|a\|^2-\rho \|b\|^2$ and Schwartz inequality, one obtains
			\begin{equation}\scalebox{0.85}{$
				\begin{aligned}
				&\mathcal{L}_x(t+T)-\mathcal{L}_x(t)\le
				-\frac{\gamma\rho}{1+\rho}\int_{t}^{t+T}\|\Sigma^{\frac 1 2}(\tau)x(t)\|^2d\tau\\
				&+\rho \gamma\lambda_{\Sigma} T
				\int_t^{t+T}\int_t^{\tau}\|f(x(s),s)\|^2ds d\tau.
				\end{aligned}$}
			\end{equation}
		Using the PE condition of matrix $\Sigma(t)$ and the fact $\dot{\mathcal{L}}_x(t)\le -\frac 1 c \|f(x,t)\|^2$, it yields
		\begin{equation}\scalebox{0.8}{$
			\begin{aligned}\label{eq:intL}
			&\mathcal{L}_x(t+T)-\mathcal{L}_x(t)\le
			-\frac{\mu T\gamma\rho}{1+\rho}\|x(t)\|^2-c\rho \gamma\lambda_{\Sigma}T
			\int_t^{t+T}\int_t^{\tau}\dot{\mathcal{L}}(s)ds d\tau.
			\end{aligned}$}
		\end{equation}
		By changing the order of integration in \eqref{eq:intL}, one gets
			\begin{equation*}\scalebox{0.85}{$
				\begin{aligned}
				\mathcal{L}_x(t+T)\le(1-\sigma)\mathcal{L}_x(t),\ \sigma:=\frac{\rho {\mu T \gamma}}{(1+\rho)(1+\rho cT^2\gamma\lambda_{\Sigma} )\lambda_2}.
				\end{aligned}$}	
			\end{equation*}
		By choosing $\rho = \frac{\lambda_2}{\mu T \gamma}$, one has $\sigma = \frac{1}{1+\rho}\frac{1}{1+ \rho c T^2\gamma\lambda_\Sigma} < 1$. For any $t\ge 0$, let $N$ be the smallest positive integer such that $t\le NT$. Since $\mathcal{L}_x(t)\leq \mathcal{L}_x((N-1)T)\leq(1-\sigma) \mathcal{L}_x((N-2)T)$, $\mathcal{L}_x(t)$ can be bounded by a staircase geometric series such that
		$\mathcal{L}_x(t) \leq (1-\sigma)^{N-1} L_x(0)$ and hence the exponential convergence follows from
		$\mathcal{L}_x(t) \leq \frac{\exp(-bNT)}{1-\sigma} \mathcal{L}_x(0)\le\frac{\exp(-bt)}{1-\sigma} \mathcal{L}_x(0) $
		with
		\scalebox{0.9}{$b = \frac{1}{T}\ln(\frac{1}{1-\sigma}) > \frac \sigma T.$}
	\end{pf}

	\begin{prop}\label{prop:rate}
		Consider the cascaded system defined in Theorem \ref{thm:fullstate}. If Assumptions 1-3 are satisfied and the convergence rate of the unforced system $\dot{\tilde x}_{ij}=-A(g_i(t))\tilde x_{ij}$ is greater than $b_{i}$, for each agent $i\in \mathcal V\backslash\{1\}$ and $j\in \mathcal N_i$. Then the convergence rate for each agent $i\in \mathcal V\backslash\{1,2\}$ of the cascaded system is greater than $c_i=\frac 1 2 \min\{c_{i-1},b_i\}$, with $c_2=b_2$, which is a lower bound obtained when the leader-follower structure is minimal and has a single directed path. Additionally, if $b_i=b$, the convergence rate for each agent $i\in \mathcal V\backslash\{1,2\}$ in the cascaded system is greater than $\frac {b}{2^{i-2}}$.
	\end{prop}
	\vspace{-0.6cm}
	\begin{pf}
	Using the same argument used in the proof of \cite[Theorem 4.9]{khalil92}
			and by mathematical induction, the convergence rate for each agent $i\in \mathcal V\backslash\{1,2\}$ in the cascaded form is greater than $c_i=\frac 1 2 \min\{c_{i-1},b_i\}$. When $b_i=b$, the conclusion follows by iterative substitution of $c_{i-1}$ in the expression for $c_i$.
	\end{pf}

\end{document}